\documentclass[10pt,twocolumn]{article} 
\usepackage{mathptmx} % This is Times font

\usepackage[normalem]{ulem}
\usepackage{algorithm}
\usepackage[noend]{algpseudocode}

\usepackage{etoolbox}
\makeatletter
\patchcmd{\ALG@doentity}{\item[]\nointerlineskip}{}{}{}
\makeatother

\usepackage{comment}
\usepackage{balance}
\usepackage{placeins}
\usepackage{caption}
\newcommand{\ignore}[1]{}
\usepackage{fancyhdr}
\usepackage[normalem]{ulem}
\usepackage[hyphens]{url}
\usepackage{hyperref}
\usepackage{subfig}
\usepackage{enumitem}
\usepackage{array}
\usepackage[table]{xcolor}
\usepackage{multirow}
\usepackage{graphicx}

\date{}

\begin{document}

\title{\vspace{-3cm} Sensitivity Analysis of Core Specialization Techniques}

\author{Prathmesh Kallurkar \footnote{The author contributed to this work while at Indian Institute of Technology Delhi} \\ Microarchitecture Research Lab \\ Intel Corporation \\ e-mail: prathmesh.kallurkar@intel.com 
\and Smruti R. Sarangi \\ Department of Computer Science \\ Indian Institute of Technology Delhi \\ e-mail: srsarangi@cse.iitd.ac.in}

%\author{Prathmesh Kallurkar, Smruti R. Sarangi}
%\author{Prathmesh Kallurkar \\ Microarchitecture Research Lab, Intel Corporation \\ prathmesh.kallurkar@intel.com
%\and Smruti R. Sarangi \\ Department of Computer Science, Indian Institute of Technology, New Delhi, India \\ srsarangi@cse.iitd.ac.in}
%\author{Prathmesh Kallurkar\Mark{1}}
%\author{Smruti R. Sarangi\Mark{2}}
%\affil[1]{\vspace{-3mm}Microarchitecture Research Lab, Intel Corporation}
%\affil{prathmesh.kallurkar@intel.com}
%\affil[2]{\vspace{-3mm}Department of Computer Science, Indian Institute of Technology, New Delhi, India}
%\affil{srsarangi@cse.iitd.ac.in}
%\affil{\vspace{-3mm}Email: \{prathmesh.kallurkar\Mark{1},srsarangi\Mark{2}\}@cse.iitd.ac.in}

\maketitle

\begin{abstract}
The instruction footprint of OS-intensive workloads such as web servers, 
database servers, and file servers typically exceeds the size of the 
instruction cache (32 KB). Consequently, such workloads incur a lot of i-cache 
misses, which reduces their performance drastically. 
Several papers~\cite{nellans_cache,flexsc,regionsched_thesis,slicc,schedtask} 
have proposed to improve the performance of such workloads 
using core specialization. In this scheme, tasks with different instruction 
footprints are executed on different cores. 
In this report, we study the performance of five 
state of the art core specialization techniques: 
{\em SelectiveOffload}~\cite{nellans_cache},
{\em FlexSC}~\cite{flexsc},
{\em DisAggregateOS}~\cite{regionsched_thesis},
{\em SLICC}~\cite{slicc},
and {\em SchedTask}~\cite{schedtask} for different system parameters.
Our studies show that for a suite of 8 popular OS-intensive workloads, 
{\em SchedTask} performs best for all evaluated configurations.
\end{abstract}

\FloatBarrier
\section{Multi-programmed Workloads}
%\section{\bf \em (A) Impact of multi-programmed workloads}

%\begin{minipage}[b]{0.3\linewidth}
%\includegraphics[width=0.97\columnwidth]{figures/graphs/kendalls_tau}
%\caption{Impact of the size of the {\em Page-heatmap register} on the quality of its ranking}
%\label{fig:kendalls_tau}
%\end{minipage}

%& 

\begin{figure}
\centering
\includegraphics[width=0.95\columnwidth]{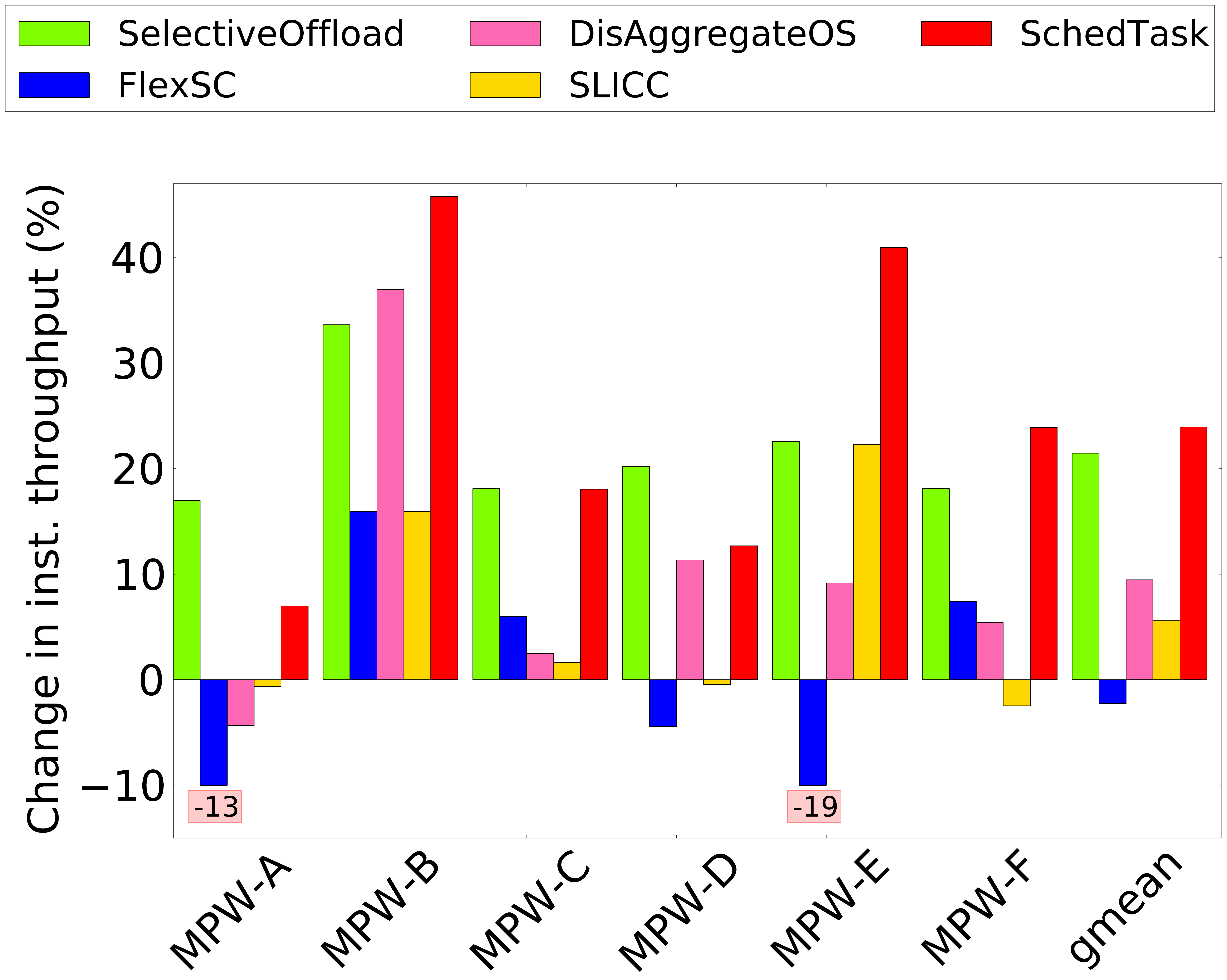}
\caption{Impact of different techniques on the instruction throughput of a system executing multi-programmed workloads}
\label{fig:mpw_performance}
\end{figure}

\begin{table}
\centering
	\scriptsize
	\begin{tabular}{|c|>{\centering \arraybackslash}p{4cm}|>{\centering \arraybackslash}p{1.5cm}|}
		\hline
		Bag ID & Constituent benchmarks  & Workload of individual benchmark\\\hline \hline
		{\em MPW-A}  & {\em DSS, FileSrv} & 1X \\ \hline
		{\em MPW-B}  & {\em Apache, OLTP} & 1X\\ \hline
		{\em MPW-C}  & {\em Apache, DSS, FileSrv, Iscp} & 0.5X \\ \hline
		{\em MPW-D}  & {\em Apache, DSS, Find, OLTP} & 0.5X\\ \hline
		{\em MPW-E}  & {\em Find, FileSrv, Iscp, Oscp}  & 0.5X\\ \hline
		{\em MPW-F}  & {\em Apache, FileSrv, MailSrvIO, OLTP} & 0.5X\\ \hline
	\end{tabular}
	\caption{Constituent benchmarks of multi-programmed workloads}
	\label{tab:mpw}
\end{table}

We compare the impact of all core specialization techniques on a server that is executing multiple 
OS-intensive applications. Table~\ref{tab:mpw} shows the constituent benchmarks and their workloads for 
each multi-programmed workload, and Figure~\ref{fig:mpw_performance} shows the impact of different 
core specialization techniques on the weighted instruction throughput of each multi-programmed workload. 
We start by allocating equal number of cores for each benchmark and then let the scheduling techniques 
decide the appropriate number of cores to execute the constituent tasks of each multi-programmed workload. 
The mean improvement in the weighted instruction throughput for these techniques is:
{\em SelectiveOffload} (21.48\%), {\em FlexSC} (-2.26\%), 
{\em DisAggregateOS} (9.47\%), {\em SLICC} (5.64\%), and {\em SchedTask} (23.94\%).
The primary point to note from Figure~\ref{fig:mpw_performance} is that 
the performance of {\em SLICC} is low for multi-programmed workloads.
This is an artifact of {\em SLICC}'s thread decomposition policy, which does not group 
common portions of OS execution across different applications. 
{\em FlexSC}, {\em DisAggregateOS}, and {\em SchedTask} group system calls based on their IDs. 
Hence, for these techniques, there is a high correlation between their performance of a multi-programmed 
workload and its constituent benchmarks. 

\begin{table*}
\centering
\scriptsize
\begin{tabular}{|>{\centering \arraybackslash}p{0.7cm}|>{\centering \arraybackslash}p{1.5cm}|c|c|c|c|c|c|c|c|c|c|c|c|c|c|c|c|c|c|}
\hline

 \multirow{2}{*}{iSize} & \multirow{2}{*}{Technique}  & \multicolumn{2}{c|}{Find} & \multicolumn{2}{c|}{Iscp} & \multicolumn{2}{c|}{Oscp} & \multicolumn{2}{c|}{Apache} & \multicolumn{2}{c|}{DSS} & \multicolumn{2}{c|}{FileSrv} & \multicolumn{2}{c|}{MailSrvIO} & \multicolumn{2}{c|}{OLTP} & \multicolumn{2}{c|}{geom. mean}\\
\cline{3-20}
 &   & iHit  & Perf  & iHit  & Perf  & iHit  & Perf  & iHit  & Perf  & iHit  & Perf  & iHit  & Perf  & iHit  & Perf  & iHit  & Perf  & iHit  & Perf \\
\hline
\multirow{5}{*}{16 KB} & SelectiveOffload & \cellcolor{lightgray}1 & 10 & \cellcolor{lightgray}1 & 21 & \cellcolor{lightgray}1 & 12 & \cellcolor{lightgray}1 & 31 & \cellcolor{lightgray}3 & 5 & \cellcolor{lightgray}0 & 6 & \cellcolor{lightgray}1 & 0 & \cellcolor{lightgray}4 & 11 & \cellcolor{lightgray}1 & 12\\
 & FlexSC & \cellcolor{lightgray}7 & -48 & \cellcolor{lightgray}6 & -40 & \cellcolor{lightgray}6 & -50 & \cellcolor{lightgray}-1 & 12 & \cellcolor{lightgray}1 & 6 & \cellcolor{lightgray}1 & 25 & \cellcolor{lightgray}2 & 12 & \cellcolor{lightgray}2 & 10 & \cellcolor{lightgray}3 & -14\\
 & DisAggregateOS & \cellcolor{lightgray}2 & 0 & \cellcolor{lightgray}1 & 14 & \cellcolor{lightgray}1 & 10 & \cellcolor{lightgray}2 & 20 & \cellcolor{lightgray}3 & 6 & \cellcolor{lightgray}1 & 16 & \cellcolor{lightgray}3 & 0 & \cellcolor{lightgray}4 & 9 & \cellcolor{lightgray}2 & 9\\
 & SLICC & \cellcolor{lightgray}1 & 4 & \cellcolor{lightgray}1 & 24 & \cellcolor{lightgray}1 & 12 & \cellcolor{lightgray}1 & 1 & \cellcolor{lightgray}2 & 5 & \cellcolor{lightgray}1 & 15 & \cellcolor{lightgray}2 & 0 & \cellcolor{lightgray}2 & 11 & \cellcolor{lightgray}1 & 8\\
 & SchedTask & \cellcolor{lightgray}2 & 11 & \cellcolor{lightgray}1 & 40 & \cellcolor{lightgray}1 & 23 & \cellcolor{lightgray}2 & 44 & \cellcolor{lightgray}2 & 10 & \cellcolor{lightgray}1 & 34 & \cellcolor{lightgray}2 & 28 & \cellcolor{lightgray}3 & 17 & \cellcolor{lightgray}1 & 25\\
\hline
\hline
\multirow{5}{*}{32 KB} & SelectiveOffload & \cellcolor{lightgray}2 & 7 & \cellcolor{lightgray}2 & 21 & \cellcolor{lightgray}1 & 8 & \cellcolor{lightgray}2 & 27 & \cellcolor{lightgray}3 & 5 & \cellcolor{lightgray}1 & 4 & \cellcolor{lightgray}3 & 0 & \cellcolor{lightgray}3 & 9 & \cellcolor{lightgray}2 & 10\\
 & FlexSC & \cellcolor{lightgray}10 & -51 & \cellcolor{lightgray}7 & -44 & \cellcolor{lightgray}6 & -56 & \cellcolor{lightgray}-1 & 7 & \cellcolor{lightgray}2 & 6 & \cellcolor{lightgray}2 & 29 & \cellcolor{lightgray}2 & 12 & \cellcolor{lightgray}1 & 4 & \cellcolor{lightgray}3 & -18\\
 & DisAggregateOS & \cellcolor{lightgray}3 & -2 & \cellcolor{lightgray}2 & 16 & \cellcolor{lightgray}1 & 4 & \cellcolor{lightgray}4 & 20 & \cellcolor{lightgray}3 & 6 & \cellcolor{lightgray}2 & 20 & \cellcolor{lightgray}5 & 4 & \cellcolor{lightgray}3 & 6 & \cellcolor{lightgray}3 & 9\\
 & SLICC & \cellcolor{lightgray}4 & 3 & \cellcolor{lightgray}2 & 28 & \cellcolor{lightgray}1 & 7 & \cellcolor{lightgray}3 & 9 & \cellcolor{lightgray}1 & 5 & \cellcolor{lightgray}1 & 20 & \cellcolor{lightgray}3 & 2 & \cellcolor{lightgray}2 & 13 & \cellcolor{lightgray}2 & 11\\
 & SchedTask & \cellcolor{lightgray}4 & 7 & \cellcolor{lightgray}3 & 39 & \cellcolor{lightgray}1 & 15 & \cellcolor{lightgray}4 & 38 & \cellcolor{lightgray}3 & 10 & \cellcolor{lightgray}2 & 44 & \cellcolor{lightgray}4 & 28 & \cellcolor{lightgray}3 & 12 & \cellcolor{lightgray}3 & 23\\
\hline
\hline
\multirow{5}{*}{64 KB} & SelectiveOffload & \cellcolor{lightgray}3 & 6 & \cellcolor{lightgray}3 & 22 & \cellcolor{lightgray}2 & 6 & \cellcolor{lightgray}4 & 26 & \cellcolor{lightgray}0 & 5 & \cellcolor{lightgray}1 & 5 & \cellcolor{lightgray}3 & -1 & \cellcolor{lightgray}2 & 8 & \cellcolor{lightgray}2 & 9\\
 & FlexSC & \cellcolor{lightgray}8 & -52 & \cellcolor{lightgray}6 & -45 & \cellcolor{lightgray}4 & -57 & \cellcolor{lightgray}1 & 8 & \cellcolor{lightgray}0 & 6 & \cellcolor{lightgray}1 & 23 & \cellcolor{lightgray}2 & 12 & \cellcolor{lightgray}1 & 4 & \cellcolor{lightgray}3 & -19\\
 & DisAggregateOS & \cellcolor{lightgray}5 & -1 & \cellcolor{lightgray}4 & 16 & \cellcolor{lightgray}2 & 3 & \cellcolor{lightgray}8 & 22 & \cellcolor{lightgray}0 & 6 & \cellcolor{lightgray}1 & 27 & \cellcolor{lightgray}4 & 5 & \cellcolor{lightgray}3 & 5 & \cellcolor{lightgray}3 & 10\\
 & SLICC & \cellcolor{lightgray}5 & 4 & \cellcolor{lightgray}3 & 33 & \cellcolor{lightgray}2 & 7 & \cellcolor{lightgray}8 & 21 & \cellcolor{lightgray}0 & 6 & \cellcolor{lightgray}1 & 26 & \cellcolor{lightgray}2 & 5 & \cellcolor{lightgray}3 & 19 & \cellcolor{lightgray}3 & 15\\
 & SchedTask & \cellcolor{lightgray}5 & 6 & \cellcolor{lightgray}4 & 39 & \cellcolor{lightgray}2 & 13 & \cellcolor{lightgray}8 & 37 & \cellcolor{lightgray}0 & 11 & \cellcolor{lightgray}1 & 36 & \cellcolor{lightgray}3 & 28 & \cellcolor{lightgray}2 & 13 & \cellcolor{lightgray}3 & 22\\
\hline
\hline

%\input{./figures/scripts/icache_table.tex}
%\multicolumn{20}{|c|}{iSize $\rightarrow$ size of the i-cache, iHit $\rightarrow$ change in i-cache hit rate (\%), Perf $\rightarrow$ change in instruction throughput (\%)} \\ \hline
\multicolumn{20}{|c|}{iSize is the size of the i-cache. iHit and Perf are the change (\%) in i-cache hit rate and the instruction throughput respectively relative to the baseline with the same i-cache size } \\ \hline
\end{tabular}
\caption{Impact of the size of the instruction cache on the instruction cache hit rate and instruction throughput}
\label{tab:appendix_icache_size}
\end{table*}

%Table~\ref{tab:expsetup} describes the configuration of the baseline system. In this appendix, we 
%study the impact of changing different system parameters on the performance of all core specialization 
%techniques. 

\section{Instruction Cache Size}
%{\bf \em (B) Impact of the instruction cache size}

Table~\ref{tab:appendix_icache_size} shows the impact of the i-cache size on the i-cache hit rate and the 
instruction throughput derived by all core specialization techniques. We evaluate all techniques 
for the following three i-cache configurations: 4-way 16 KB, 4-way 32 KB, and 4-way 64 KB. 
A baseline system with a smaller i-cache incurs more cache misses and therefore, 
the core specialization techniques can improve instruction throughput better. 
Our proposed technique 
improves throughput by 25\%, 23\%, and 22\% over the baseline for a 16 KB, 32 KB, and a 64 KB i-cache system,
respectively. This results in a performance improvement of 13\%, 12\%, and 7\% respectively over
the best state of the art techniques.

\begin{table*}
\centering
\scriptsize
\begin{tabular}{|>{\centering \arraybackslash}p{2cm}|>{\centering \arraybackslash}p{1.5cm}|c|c|c|c|c|c|c|c|c|}
\hline
\multirow{2}{*}{Cache configuration} & \multirow{2}{*}{Technique}  & Find & Iscp & Oscp & Apache & DSS & FileSrv & MailSrvIO & OLTP & geom. mean\\
\cline{3-11}
 & & \multicolumn{9}{c|}{Change in the instruction throughput (\%) relative to the baseline system with the same cache configuration} \\ \hline
\hline
\multirow{5}{*}{Config1} & SelectiveOffload & -1 & 18 & 14 & 18 & 9 & 13 & 17 & 12 & 12\\
 & FlexSC & -57 & -51 & -60 & 17 & 1 & 11 & 20 & 8 & -21\\
 & DisAggregateOS & -7 & 9 & 10 & 0 & 2 & 16 & 25 & 9 & 7\\
 & SLICC & 3 & 27 & 16 & 20 & 6 & 18 & 15 & 18 & 15\\
 & SchedTask & 11 & 36 & 21 & 38 & 2 & 30 & 33 & 14 & 23\\
\hline
\hline
\multirow{5}{*}{Config2} & SelectiveOffload & -2 & 16 & 14 & 16 & 10 & 10 & 13 & 10 & 11\\
 & FlexSC & -59 & -53 & -61 & 15 & 1 & 12 & 19 & 5 & -23\\
 & DisAggregateOS & -9 & 1 & 0 & 18 & 2 & 10 & 21 & 8 & 6\\
 & SLICC & 1 & 25 & 16 & 18 & 5 & 18 & 11 & 15 & 13\\
 & SchedTask & 7 & 33 & 20 & 31 & 2 & 27 & 28 & 10 & 19\\
\hline
\hline
\multirow{5}{*}{Config 3} & SelectiveOffload & 7 & 21 & 8 & 27 & 5 & 4 & 0 & 9 & 10\\
 & FlexSC & -51 & -44 & -56 & 7 & 6 & 29 & 12 & 4 & -18\\
 & DisAggregateOS & -2 & 16 & 4 & 20 & 6 & 20 & 4 & 6 & 9\\
 & SLICC & 3 & 28 & 7 & 9 & 5 & 20 & 2 & 13 & 11\\
 & SchedTask & 7 & 39 & 15 & 38 & 10 & 44 & 28 & 12 & 23\\
\hline
\hline

\multicolumn{1}{|c}{\multirow{2}{*}{Config1 \hspace{0.2cm} $\rightarrow$}} & \multicolumn{10}{l|}{{\em Private caches} \hspace{0.4cm} i-cache and d-cache: (4-way 32 KB. latency = 3 cycles)} \\ 
\multicolumn{1}{|c}{} & \multicolumn{10}{l|}{{\em Shared cache} \hspace{0.5cm} L2 cache: (8-way 8 MB. latency = 18 cycles)} \\ 
\hline

\multicolumn{1}{|c}{\multirow{2}{*}{Config2 \hspace{0.2cm} $\rightarrow$}} & \multicolumn{10}{l|}{{\em Private caches} \hspace{0.4cm} i-cache and d-cache: (4-way 32 KB. latency = 3 cycles)} \\ 
\multicolumn{1}{|c}{} & \multicolumn{10}{l|}{{\em Shared cache} \hspace{0.5cm} L2 cache: (8-way 8 MB. latency = 8 cycles)} \\ 
\hline

\multicolumn{1}{|c}{\multirow{2}{*}{Config3 \hspace{0.2cm} $\rightarrow$}} & \multicolumn{10}{l|}{{\em Private caches} \hspace{0.4cm} i-cache and d-cache: (4-way 32 KB. latency = 3 cycles), \hspace{0.2cm} L2 cache: (4-way 256 KB. latency = 8 cycles)} \\ 
\multicolumn{1}{|c}{} & \multicolumn{10}{l|}{{\em Shared cache} \hspace{0.5cm} L3 cache: (8-way 8 MB. latency = 18 cycles)} \\ 
\hline

\end{tabular}
\caption{Impact of the cache configuration on the instruction throughput}
\label{tab:appendix_cache_configuration}
\end{table*}

\section{Cache Configuration}
%{\bf \em (C) Impact of the cache configuration}

Table~\ref{tab:appendix_cache_configuration} describes three cache configurations (Config1, Config2, and Config3) 
and their impact on the instruction throughput of all techniques. Config1 and Config2 have two levels of cache 
hierarchy whereas Config3 has three levels of cache hierarchy. Since the performance benefit derived by a 
core specialization technique is directly proportional to the i-cache miss penalty, the performance of all 
techniques is the least for Config2 and the most for Config1. 
Our proposed technique 
improves throughput by 24\%, 21\%, and 23\% over the baseline for a system with Config1, Config2, and Config3 
cache configurations respectively. 
This results in a 7, 6, and 12 percentage point enhancement in performance (respectively)
over the best existing techniques.

%Since a baseline system with an i-cache size of 16 KB has a very low i-cache hit rate (75-85\%), 
%such a system benefits the most with a core specialization technique. Consequently, among all evaluated 
%i-cache configurations, we observe that the performance benefits derived for all the techniques 
%are the highest for a 16 KB i-cache system. 
%Addtionally we note that the SLICC technique gets fewer opportunities to partition tasks with a 16 KB instruction 
%footprint granularity.  Consequently, for a 16 KB i-cache system, the Nellans technique outperforms the SLICC 
%technique by 1\% (mean).  
%
%{\em \bf 32 KB: } The results for a 32 KB i-cache configuration are the same as described in the Section 5.1 of 
%the submitted paper. 
%
%{\em \bf 64 KB: } A 64 KB cache is sufficient to hide almost all i-cache misses for the non OS-intensive 
%workloads such as the SPEC CPU2006, Parsec, and Splash2 benchmark suites. However, since OS-intensive 
%workloads have large instruction footprints (150-200 KB), the i-cache hit rates of such workloads with 
%a baseline Linux system is still around 88-96\%. Hence, there is an opportunity to improve the performance 
%of such workloads with core specialization techniques. As shown in Table~\ref{tab:appendix_icache_size}
%SLICC and SchedTask are the best performing techniques for a 64 KB i-cache configuration.

\begin{table*}
\centering
\scriptsize
\begin{tabular}{|>{\centering \arraybackslash}p{0.8cm}|>{\centering \arraybackslash}p{1.5cm}|c|c|c|c|c|c|c|c|c|}
\hline

\multirow{2}{*}{\#cores} & \multirow{2}{*}{Technique}  & Find & Iscp & Oscp & Apache & DSS & FileSrv & MailSrvIO & OLTP & geom. mean\\
\cline{3-11}
 & & \multicolumn{9}{c|}{Change in the instruction throughput (\%) relative to the baseline system with the same number of cores} \\ \hline
\hline
\multirow{5}{*}{8 cores} & SelectiveOffload & 14 & 22 & 17 & 48 & 5 & 2 & -1 & 17 & 15\\
 & FlexSC & -24 & -26 & -41 & 13 & 6 & 5 & 12 & 3 & -8\\
 & DisAggregateOS & -17 & -14 & -16 & 0 & -10 & -19 & -28 & -1 & -14\\
 & SLICC & 6 & -5 & -13 & -4 & -3 & -10 & -11 & -5 & -6\\
 & SchedTask & 20 & 24 & 10 & 36 & 9 & 16 & 22 & 12 & 18\\
\hline
\hline
\multirow{5}{*}{16 cores} & SelectiveOffload & 19 & 27 & 26 & 47 & 4 & 6 & 0 & 23 & 18\\
 & FlexSC & -24 & -24 & -40 & 13 & 5 & 4 & 15 & 13 & -7\\
 & DisAggregateOS & -1 & -10 & -14 & 3 & -5 & -15 & -23 & 4 & -8\\
 & SLICC & 17 & 6 & -2 & 3 & 3 & -3 & -4 & 8 & 3\\
 & SchedTask & 32 & 37 & 22 & 51 & 8 & 17 & 31 & 26 & 27\\
\hline
\hline
\multirow{5}{*}{24 cores} & SelectiveOffload & 15 & 29 & 16 & 40 & 5 & 6 & 0 & 15 & 15\\
 & FlexSC & -45 & -35 & -53 & 11 & 8 & 22 & 13 & 10 & -13\\
 & DisAggregateOS & -4 & 1 & -1 & 6 & 0 & 2 & -12 & 4 & 0\\
 & SLICC & 7 & 25 & 6 & 6 & 8 & 9 & 0 & 13 & 9\\
 & SchedTask & 15 & 47 & 23 & 51 & 13 & 27 & 28 & 18 & 27\\
\hline
\hline
\multirow{5}{*}{32 cores} & SelectiveOffload & 7 & 21 & 8 & 27 & 5 & 4 & 0 & 9 & 10\\
 & FlexSC & -51 & -44 & -56 & 7 & 6 & 29 & 12 & 4 & -18\\
 & DisAggregateOS & -2 & 16 & 4 & 20 & 6 & 20 & 4 & 6 & 9\\
 & SLICC & 3 & 28 & 7 & 9 & 5 & 20 & 2 & 13 & 11\\
 & SchedTask & 7 & 39 & 15 & 38 & 10 & 44 & 28 & 12 & 23\\
\hline
\hline

%\input{./figures/scripts/numcores_table.tex}
%\multicolumn{20}{|c|}{Idle $\rightarrow$ fraction of idle time (\%), Perf $\rightarrow$ change in instruction throughput (\%)} \\ \hline
\end{tabular}
\caption{Impact of the number of cores on the instruction throughput}
\label{tab:appendix_numcores}
\end{table*}

\section{Number of Cores}
%{\bf \em (D) Impact of the number of cores}

Table~\ref{tab:appendix_numcores} shows the impact of the number of cores on the instruction throughput 
of different core specialization techniques. We evaluate all the techniques for the following four systems: 
system with 8 cores, system with 16 cores, system with 24 cores, and a system with 32 cores. We do not 
consider a system with less than 8 cores because such a system is not practical for the OS-intensive server-class 
workloads that we consider. Our proposed technique improves throughput by 18\%, 27\%, 27\%, and 23\% 
over the baseline for a system with 8 cores, 16 cores, 24 cores, and 32 cores respectively.
This results in 3, 9, 12, and 12 percentage points enhancements, respectively, over the best existing techniques.

\section{Instruction Prefetcher}
%{\bf \em (E) Impact of instruction prefetcher}

\begin{figure}
\centering
\includegraphics[width=0.95\columnwidth]{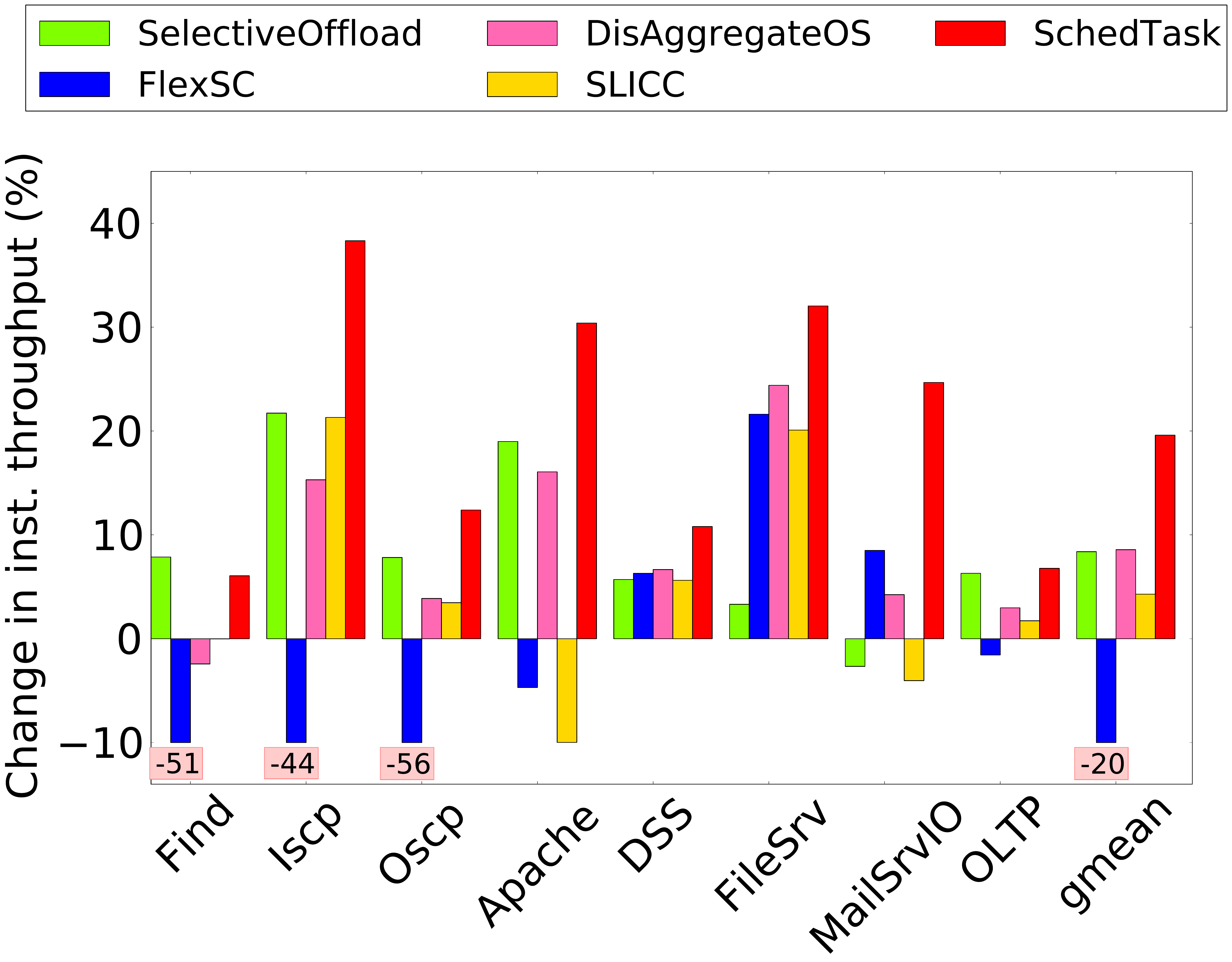}
\caption{Impact of the instruction prefetcher on the instruction throughput}
\label{fig:cgp_performance}
\end{figure}

Figure~\ref{fig:cgp_performance} shows the impact of core specialization techniques on the
instruction throughput when the baseline system employs a hardware instruction prefetcher.  We
use the hardware-only mode ({\tt CGHC-2K+32K}) of the Call Graph Prefetcher (CGP)~\cite{CGP} as the
instruction prefetcher. We use CGP because its hardware overheads are not very high and it is shown
to give better performance than the classical instruction prefetchers such as next-line prefetcher
and correlation-based prefetcher~\cite{markov_nesbit}.  We observe that CGP reduces the number of i-cache
misses by 20-30\% and thus improves the performance of a system without an  instruction prefetcher
by around 4-5\%\footnote{The original paper~\cite{CGP} uses a 2-level memory hierarchy only and hence it 
enhances performance more}. Since a baseline system with CGP incurs fewer i-cache misses, the scheduling
techniques gain lesser by improving the instruction locality. The mean improvements in the
instruction throughput of the system after employing CGP are: {\em SelectiveOffload} (8.37\%), {\em
FlexSC} (-20.93\%), {\em DisAggregateOS} (8.57\%), {\em SLICC} (4.28\%), and {\em SchedTask}
(19.6\%). 

\section{Trace Cache}
\begin{figure}
\centering
\includegraphics[width=0.95\columnwidth]{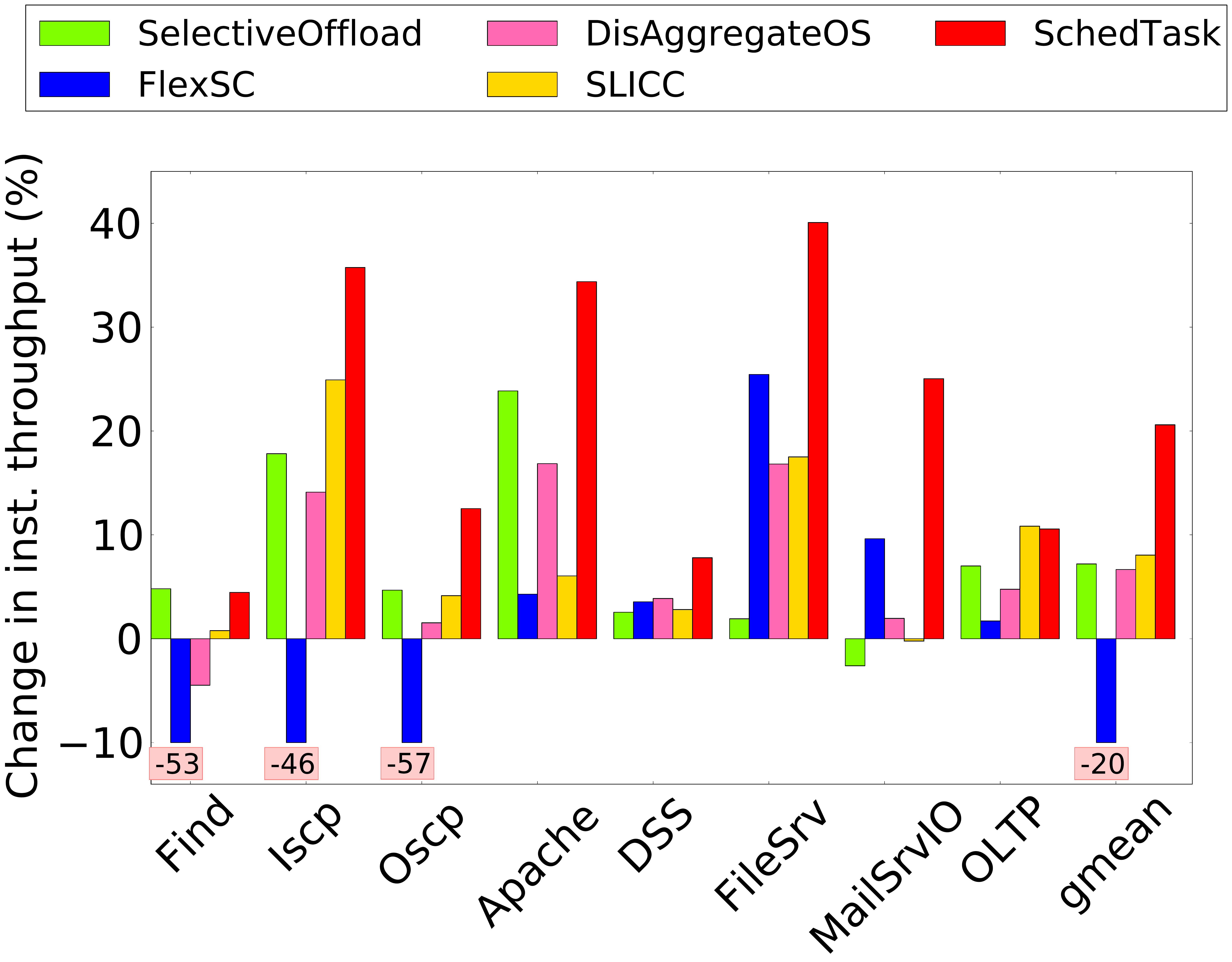}
\caption{Impact of the trace cache on the instruction throughput}
\label{fig:tc_performance}
\end{figure}

%{\bf \em (F) Impact of trace cache}
Figure~\ref{fig:tc_performance} shows the impact of core specialization
techniques on the instruction throughput when the baseline system employs a
trace cache. We use the trace cache implementation that was proposed in
~\cite{tracecache}. We observe that since the instruction footprints of the
considered workloads are very large ($>$250KB), traces belonging to different
{\em SuperFunctions} keep evicting each other from the shared trace cache.
Hence, the performance gains derived by using core specialization techniques
do not change much for a system employing a trace cache versus one that
does not employ a trace cache. For a system that employs a trace cache, the mean
performance gains derived by different techniques are: {\em SelectiveOffload}
(7.2\%), {\em FlexSC} (-20.38\%), {\em DisAggregateOS} (6.67\%), {\em SLICC}
(8.04\%), and {\em SchedTask} (20.6\%). 

\section{Conclusion}
In this report, we studied the sensitivity of 
five state of the art core specialization 
techniques to multi-programmed workloads, cache configurations, 
instruction prefetchers, and trace-cache. 
Our studies show that {\em SchedTask}~\cite{schedtask} 
outperforms other techniques~\cite{nellans_cache,flexsc,regionsched_thesis,slicc} 
for all evaluated configurations. 
This is because {\em SchedTask} employs a fine-grained task 
scheduler and a superior work stealing algorithm.

\section*{Acknowledgment}
We thank Omais Pandith and Himani Raina 
for providing us their Tejas model of 
``Trace Based Instruction Caching''; it helped us 
evaluate the impact of Trace Caches on different 
core specialization techniques.

\FloatBarrier
\pagebreak
\balance

\bibliographystyle{abbrv}
\bibliography{references}

\end{document}